\documentclass[twocolumn,showpacs,preprintnumbers,amsmath,amssymb,floatfix]{revtex4}
\usepackage{graphicx}

\begin{document}


\title{Zero-bias anomalies in electrochemically fabricated nanojunctions}

\author{L.H. Yu and D. Natelson}

\affiliation{Department of Physics and Astronomy, Rice University, 6100 Main St., Houston, TX 77005}

\date{\today}

\begin{abstract}

A streamlined technique for the electrochemical fabrication of metal
nanojunctions (MNJs) between lithographically defined electrodes is
presented.  The first low-temperature transport measurements in such
structures reveal suppression of the conductance near zero-bias.  The
size of the zero-bias anomaly (ZBA) depends strongly on the
fabrication electrochemistry and the dimensions of the resulting MNJ.
We present evidence that the nonperturbative ZBA in atomic-scale
junctions is due to a density of states suppression in the leads.

\end{abstract}

\maketitle


Metallic nanojunctions (MNJ) are unique tools for examining electronic
transport, correlations, quantum coherence, and disorder at the atomic
scale.  These atomic size devices, with diameter and length much
smaller than the electronic phase coherence length, are the subjects
of current research.  Extensive progress has been made in
understanding {\it clean} MNJs fabricated and measured in ultrahigh
vacuum (UHV) by the break junction method, using either scanning
tunneling microscope\cite{GimzewskietAl87PRB} or microfabricated
structures\cite{RuitenbeeketAl96RSI,Ruitenbeek00}.

Recently MNJs have also been prepared using electrochemistry and
proposed for use in molecular electronic
devices\cite{MorpurgoetAl99APL,LietAl00APL,WuetAl00IEEE,BoussaadetAl02APL,KervennicetAl02APL}.
However, MNJs made in solution may be ``dirty" due to grain boundaries
and incorporation or surface adsorption of ionic impurities.  While
junctions made electrochemically exhibit conductance quantization and
Ohmic behavior at room
temperature\cite{MorpurgoetAl99APL,LietAl00APL}, low-temperature
transport properties of these systems have yet to be examined.

In this letter we report a nanojunction fabrication method that
synthesizes elements from various electrochemical approaches, and
report low-temperature nanojunction conductance measurements.  We find
suppressions of the dc conductance near zero bias at low temperatures.
In Au junctions made using an alkaline buffer
solution\cite{MorpurgoetAl99APL} and having
conductance $G(300~{\rm K})< \sim 2e^{2}/h \equiv G_{0}$, this
suppression approaches 100\%.  Much smaller zero-bias anomalies (ZBAs)
occur in Au junctions made with an HCl solution\cite{BoussaadetAl02APL}.
We show that the large ZBAs are caused by nonperturbative corrections
to the local density of states of the leads.

\begin{figure}[h!]
\begin{center}
\includegraphics[clip, width=7.5cm]{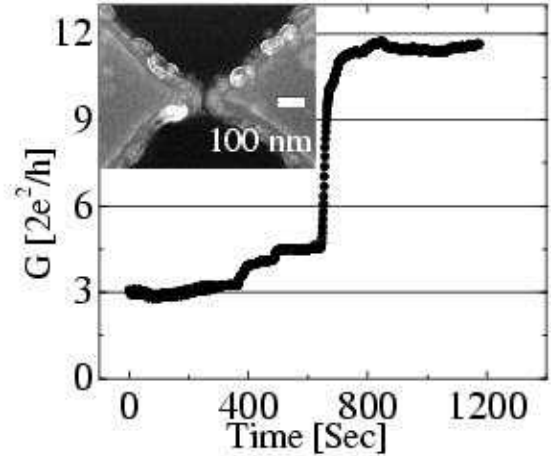}
\end{center}
\vspace{-3mm}
\caption{\footnotesize Steps in junction conductance as
atomic configurations shift during growth.  Inset: micrograph
of alumina-coated starting electrodes.}
\label{fig:device}
\end{figure}

The nanojunction fabrication process begins with gold source and drain
electrodes defined by electron-beam lithography (EBL) on 200~nm of
SiO$_{2}$ on $p$+ silicon substrates.  We define electrodes with tip
widths $\sim$~200~nm separated by $\sim$10~nm.  In an e-beam
evaporator with base pressure of 10$^{-7}$~mB, we then deposit 2.5~nm
of titanium (0.1 nm/s), 25~nm of gold (0.3 nm/s), and 20~nm of
Al$_{2}$O$_{3}$ (0.3 nm/s), and perform liftoff.  The electrodes have
$R/\Box \sim 1~\Omega$.  A representative image of an electrode pair
is in Fig.~\ref{fig:device}.  An additional 0.1~mm$^{2}$ pad 20~$\mu$m
from the tips of the electrodes is also evaporated to serve as an
auxiliary electrode\cite{WuetAl00IEEE}.

Starting from these electrodes, electrochemistry is performed to
produce a nanoscale junction.  The low-temperature junction
conductance properties are strongly influenced by the electrochemical
protocol.  Except where noted, the devices described in this letter
are made using the following solution-based
technique\cite{MorpurgoetAl99APL} (labeled ``M''): The electrodes are
covered with an aqueous solution of 0.01~M KAu(CN)$_{2}$ in a buffer
of 1~M KHCO$_{3}$ and 0.2~M KOH (pH $\approx$ 10).  Deposition current
(up to 4~$\mu$A) is sourced via the auxiliary electrode, while
maintaining the electrode pair at relative ground.  The interelectrode
conductance is measured {\it in situ} using a lock-in
amplifier\cite{MorpurgoetAl99APL}.  The alumina layer limits gold
deposition to the edges of the electrodes, and mitigates background
solution conduction.  An analogous layer requiring plasma deposition
and a focused ion beam has been used in other junction
experiments\cite{LietAl00APL}.  Evaporating the dielectric layer
yields a similar result with substantially fewer steps.  Average
resistivity of the electrodeposited gold at the electrode edges at
4.2~K is approximately 20~$\mu\Omega$-cm.

\begin{figure}[h!]
\begin{center}
\includegraphics[clip, width=7.5cm]{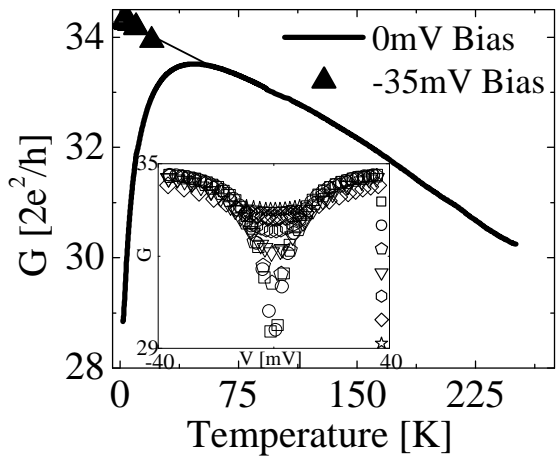}
\end{center}
\vspace{-3mm}
\caption{\footnotesize Temperature sweeps of $dI/dV$ at low and high dc bias
in a relatively large junction; Inset: bias sweeps at various $T$
showing zero-bias anomaly.  Symbols top to bottom are for $T=$1.8~K, 2.5~K, 6~K, 10~K, 15~K, 20~K, and 25~K. }
\label{fig:zbasmall}
\end{figure}

When an atomic-scale connection is formed initially between 
the electrodes, discrete conductance steps on the order of $G_{0}$
are observed (Fig.~\ref{fig:device}).  By adjusting the deposition
current, we stop the growth near a specific junction conductance.  The
final step is the removal of the solution using dry nitrogen gas.

At room temperature junctions of a few $G_{0}$ are
stable as long as tens of minutes, longer than those made in
UHV\cite{RuitenbeeketAl96RSI,Ruitenbeek00,Stafford02PSSB}, as expected
in ``dirty" junctions\cite{HansenetAl00APL}.  Discrete spontaneous
conductance changes are frequently observed, including breakage and
coalescence.  Such changes are much more rare at lower
temperatures, consistent with junctions made from a small number of
gold atoms that can diffuse at room temperature.

We have also fabricated gold nanojunctions by an alternative
electrochemical method\cite{BoussaadetAl02APL} (labeled ``BT''), using
a 0.2~M HCl solution and applying a bias voltage between the source
and drain.  By varying a series resistor, nanojunctions with various
conductance values resulted.  The stability of these nanojunctions is
comparable to those described above.

\begin{figure}[h!]
\begin{center}
\includegraphics[clip, width=7.5cm]{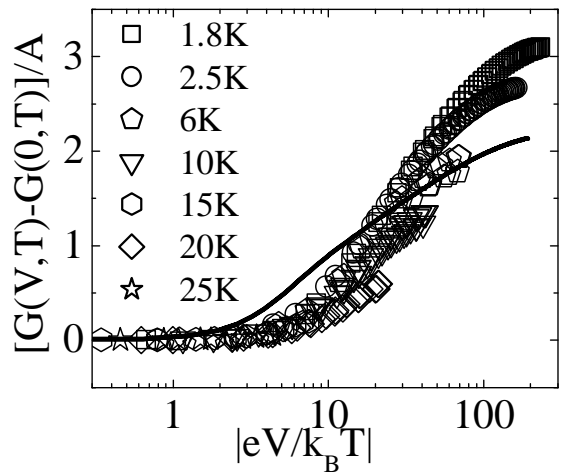}
\end{center}
\vspace{-3mm}
\caption{\footnotesize Scaled ZBA data, showing collapse to one curve at low
values of $T$, $V$, similar to that in
Ref.~\protect{\cite{WeberetAl01PRB}}.  Solid line shows best fit to
the treatment of Ref.~\protect{\cite{GolubevetAl01PRL}}.}
\label{fig:zbascale}
\end{figure}

We have measured ten surviving nanojunctions with room temperature
conductances ranging from 0.5 to 200~$G_{0}$.  Thermal contraction
effects upon cooling appear to be negligible, except as discussed
below.  We measure differential conductance $G(V,T)\equiv dI/dV$, as a
function of temperature and dc bias voltage using a quasi-4 terminal
lock-in method to eliminate the resistance of the cryostat leads.  All
nanojunctions show Ohmic behavior up to 200~mV at 300~K.  Samples with
$G(T=300$~K$) >$ 100~$G_{0}$ exhibit metallic behavior from 300~K to
1.8~K.  In smaller nanojunctions, however, we find a suppression of
the zero-bias conductance below 50~K.  This suppression, shown in
Fig.~\ref{fig:zbasmall}, is logarithmic in $T$.  Metallic behavior in
$G(V,T)$ is recovered when a -35~mV dc bias is applied.  We estimate
the cross-section $A$ of this junction to be 3~nm$^{2}$ by a Sharvin
relation\cite{Sharvin65JETP,MureketAl93PRL} for a ballistic point
contact ($w,L << \ell$, the elastic mean free path), $G/G_{0}= A
k_{\rm F}^{2}/4\pi$, and $k_{\rm F}=1.2 \times 10^{10}$~m$^{-1}$.  The
inset shows $G(V,T)$ for this junction below 30~K.  The suppression
becomes more pronounced as $T \rightarrow 0$, reaching about 15\% at
1.8~K.  A substantial low temperature ZBA is observed in all samples
with $G(300~{\rm K}) < \sim 30 G_{0}$ made using the M
electrochemistry.  The ZBAs in the BT junctions are much smaller,
about 0.2\% at 1.8~K.  The application of a constant 8.5 Tesla
magnetic field perpendicular to the nanojunction has no detectable
effect on the ZBA.

To understand the ZBA we observe in Fig.~\ref{fig:zbasmall}, we
consider the analysis in Ref.~\cite{WeberetAl01PRB} of electronic
transport in larger gold quasi-2d diffusive nanobridges.  The ZBAs in
such nanobridges are ascribed\cite{WeberetAl01PRB} to perturbative
electron-electron interaction corrections to the density of states
(DOS)\cite{AltshuleretAl85} in those devices.  At finite temperature
Weber {\it et al.} predict the conductance of the nanobridge to scale
as $[G(V,T)-G(0,T)]/A = f(eV/k_{\rm B}T)$, where $A$ describes the
temperature dependence of the zero bias conductance, $(G(0,T) = G_{*}
+ A \ln(T))$ and $f(x)$ is a function determined by the diffusive
nature of the nanobridge.  Figure \ref{fig:zbascale} shows our data
scaled this way.  At $|eV/k_{\rm B}T|<1$, $G$ is essentially constant,
corresponding to thermal smearing.  At $|eV/k_{\rm B}T|>1$, $G$ is
roughly logarithmic, qualitatively consistent with the data of Weber
et al.  We do not expect quantitative agreement, since our
nanojunction is nondiffusive.  A treatment of interaction-based ZBAs
in $G >> G_{0}$ coherent systems\cite{GolubevetAl01PRL} fits well with
$G(0,T)$, but deviates systematically from the data at finite bias, a
trend under investigation.

\begin{figure}[h!]
\begin{center}
\includegraphics[clip, width=7.5cm]{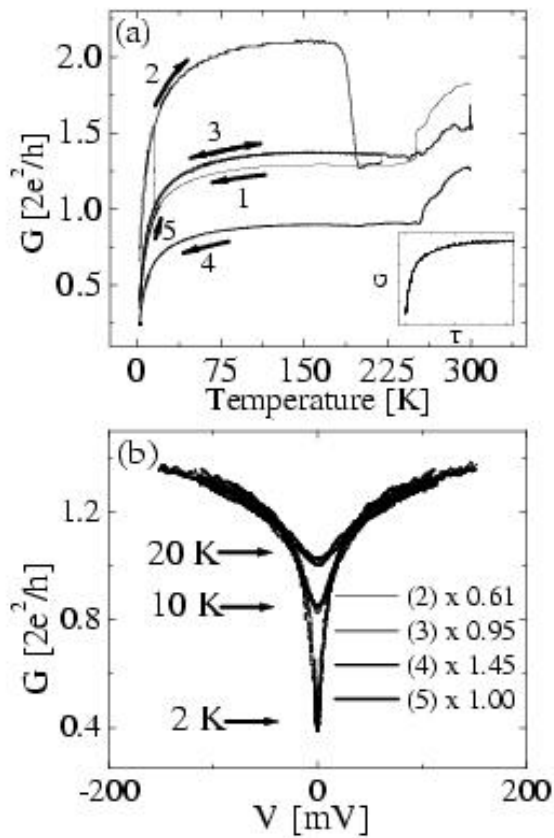}
\end{center}
\vspace{-3mm}
\caption{\footnotesize (a) Temperature cycling of a $G(300~{\rm K})\approx
1~G_{0}$ junction, showing discrete, hysteretic behavior; inset: all
the branches of $G(0,T)$ collapse onto a single curve when each is
scaled by a multiplicative constant; (b) ZBA data from the junction
configurations of (a), all collapsed by scaling factors obtained from
$G(V=0,T)$. }
\label{fig:zbabig}
\end{figure}

Atomic-scale junctions with correspondingly lower values of
$G(300~{\rm K})$ exhibit larger ZBAs at low temperatures.  Figure
\ref{fig:zbabig} shows $G$ of one informative atomic-scale
nanojunction $(G(T=300~{\rm K}) \approx 1~G_{0}$) as a function of
temperature cycling.  During the initial cool down (1), the
$G(0,T)$ is nearly $T$-independent until below 50~K, when a
substantial drop appears as in previous samples.  Upon cooling to
15~K, $G$ spontaneously increases by about 0.6~$G_{0}$.  As $T$ is
further reduced, $G$ continues to decrease, but from a higher baseline
conductance.  When the nanojunction is warmed (2), it appears to have
a high temperature $G \approx 2~G_{0}$.  However, at 220~K the
conductance of the nanojunction spontaneously decreases by about
0.85~$G_{0}$, returning near its original value.  Additional cycling
and exposure to light produce further branches of $G(V=0,T)$ (3,4,5).
Remarkably, all these $G(V=0,T)$ curves collapse onto {\it one curve}
(see inset) when each branch of $G(T)$ is multiplied by a constant.

The discrete changes confirm the atomic scale of the nanojunction.
The addition of a single partially transmitting channel as $T
\rightarrow 0$ presumably occurs as thermal contraction brings the
source and drain closer together.  The loss of the channel on warming
supports this view, with thermal expansion now stretching the
junction.  This sort of hysteresis behavior is seen in break junction
measurements.

For each conductance branch described above, the ZBA is measured at
several temperatures.  Figure \ref{fig:zbabig}b shows $G(V,T)$ at 2~K,
10~K, and 20~K for this nanojunction in its various configurations.
Scaling by the factors used to collapse the $G(V=0,T)$ branches, the
ZBA data also collapse.  Clearly $G(V,T)$ is of the form $B \times
g(V,T)$, where $B$ is a constant characteristic of a particular
configuration of junction atoms, and $g(V,T)$ is a single function
applicable to all the conductance branches.  We hypothesize that the
size and form of $g(V,T)$ is related to impurities and disorder in the
leads, and that such disorder is fabrication method- and
sample-dependent.  Note that $\Delta G/G$ for this ZBA is nearly 70\%;
we find similar nonperturbative ($>$ 60\%) ZBAs in three other M
samples with similar values of $G(300~{\rm K})$.  Such large 
suppressions are never seen in higher conductance junctions.

A natural interpretation of the scaling described above is that these
nanojunctions probe the local DOS of the leads.  The differential
conductance measurements are then analogous to the tunneling
conductance experiments used to confirm the perturbative DOS
anomalies\cite{AltshuleretAl85} in planar tunnel junctions.  The
transmittance of the junction is approximately independent of
temperature and energy, and depends on the precise configuration of
two or three mobile atoms.  The DOS at the junction-lead boundary,
however, is temperature and energy dependent, and is determined by
disorder ``built in'' during the electrodeposition process.  It is
this DOS that develops a nonperturbative suppression at low
temperatures.  A further discussion of this will be presented
elsewhere\cite{YuetAl03}.

A simple and reliable electrochemical method is used to fabricate
atomic size nanojunctions with room temperature conductance ranging
from 0.5~$G_{0}$ to 200~$G_{0}$.  The nanojunctions are stable in
ambient environment for tens of minutes and indefinitely below 200~K
under vacuum.  Differential conductance measurements while varying $T$
and $V$ reveal zero-bias anomalies at low temperatures.  Scaling of
data from multiple configurations of a particular sample shows that
large ZBAs in atomic-scale junctions are due to nonperturbative
corrections to the local density of states in the disordered metal
leads near the junctions.  The disorder in the leads is presumably due
to the particular solution-based electrochemistry used.  These
nontrivial DOS properties imply that electrochemically fabricated
nanoelectrodes may be poor choices for use in molecular electronics
experiments.

The authors gratefully acknowledge the support of the Robert A. Welch
Foundation and the Research Corporation.



\end{document}